\documentstyle[11pt,newpasp,twoside, epsf]{article}
\def\edcomment#1{\iffalse\marginpar{\raggedright\sl#1\/}\else\relax\fi}
\reversemarginpar

\begin{document}
\title{What does the Unexpected Detection of Water Vapor in Arcturus' Atmosphere Tell us?}
 \author{Nils Ryde, David L. Lambert}
  \affil{McDonald Observatory and Department of
Astronomy\\The University of Texas at Austin\\RLM 15.308\\Austin,
TX 78712}

\author{Matthew J. Richter\footnote{Visiting Astronomer at the Infrared Telescope Facility which is operated by the University  of Hawaii
under contract to the National Aeronautics and Space
Administration.}}
\affil{Department of Physics\\ The University of California, Davis\\
Davis, CA 95616}

\author{John H. Lacy$^1$, Thomas K. Greathouse$^1$}
\affil{McDonald Observatory and Department of Astronomy\\The
University of Texas at Austin\\RLM 15.308\\Austin, TX 78712}

\begin{abstract}
In this talk I presented and discussed our unexpected detection of
water vapor in the disk-averaged spectrum of the K2IIIp red giant
Arcturus [for details, see Ryde et al. (2002)]. Arcturus, or
$\alpha$ Bo\"otes is, with its effective temperature of $4300$ K,
the hottest star yet to show water vapor features. We argue that
the water vapor is photospheric and that its detection provides us
with new insights into the outer parts of the photosphere. We are
not able to model the vater vapor with a standard, one-component,
1D, radiative-equilibrium, LTE model photosphere, which probably
means we are lacking essential physics in such models. However, we
are able to model several OH lines of different excitation and the
water-vapor lines satisfactorily after lowering the temperature
structure of the very outer parts of the photosphere at $\log
\tau_{500}=-3.8$ and beyond compared to a flux-constant,
hydrostatic, standard {\sc marcs} model photosphere. Our new
semi-empirical model is consistently calculated from the given
temperature structure. I will discuss some possible reasons for a
temperature decrease in the outer-most parts of the photosphere
and the assumed break-down of the assumptions made in classical
model-atmosphere codes.  In order to understand the outer
photospheres of these objects properly, we will, most likely, need
3D hydrodynamical models of red giants also taking into account
full non-LTE and including time-dependent effects of, for example,
acoustic wave heating sensitive to thermal instabilities.
\end{abstract}

\section{OBSERVATIONS}
The data we analysed were obtained with the TEXES (Texas Echelon
Cross-echelle Spectrograph; see Lacy et al. 2002) mounted on the
NASA InfraRed Telescope Facility on Mauna Kea. TEXES is capable of
recording truly unique mid-infrared spectra. We used the
spectrograph in its high-resolution mode, with $R\sim 80,000$
covering the regions 806-822 cm$^{-1}$ (12.2-12.4 $\mu$m) and
884-925 cm$^{-1}$ (10.8-11.3 $\mu$m). TEXES is actually a
prototype for the EXES spectrometer, which is a PI instrument to
be used on the American-German Stratospheric Observatory for
Infrared Astronomy (SOFIA) which is scheduled to fly in 2004. One
of the largest advantages of flying an observatory above the
tropopause at an altitude of 12-14 km compared to a ground-based
one, is the smaller interference of telluric water vapor which
absorbes heavily in the thermal infrared. SOFIA will be flying
above 99\% of the telluric water vapor. EXES has a 1 meter long
echelon grating and TEXES has a 90 cm grating. The detectors are
Si:As $265 \times 265$ pixels detectors, which are sensitive to
mid-infrared light from ca. 4.5 to 28.5 microns.

\section{MODELLING}

We identified the unexpected absorption features in the
mid-infrared spectrum of Arcturus as water vapor and OH lines by
comparing with sun-spot spectra. Since there does not exist any
water-vapor line-lists with wavelengths accurate enough for our
high-resolution work, we constructed a new one with wavelengths
from recent laboratory measurements by Polyansky et al. (1996,
1997a, \& 1997b). The lines studied here are pure rotational
transitions of OH of the three first vibrational levels and pure
rotational transition of water within the ground vibrational
state. Vibration-rotation lines of water vapor lie in the near to
mid infrared region and electronic transitions in the UV.

We model our observations with the {\sc marcs} model atmosphere
code. These 1D hydrostatic, spherical model photospheres are
computed on the assumptions of Local Thermodynamic Equilibrium
(LTE), homogeneity and the conservation of the total flux
(radiative plus convective; the convective flux being computed
using the mixing length formulation). The radiative field used in
the model generation, is calculated with absorption from atoms and
molecules by opacity sampling in approximately 84\,000
wavelength points over the entire, relevant wavelength range
considered for the star ($2300\,\mbox{\AA} $--$
20\,\mbox{$\mu$m}$). The fundamental parameters for our standard
model are $T_{\rm eff}= 4300$ K, $\log g = 1.5$ (cgs),
[Fe/H]$=-0.5$, and $\xi_t=1.7$ km s$^{-1}$, and use the abundances
of carbon, nitrogen, and oxygen given by Decin et al. (1997).
Synthetic spectra are subsequently calculated.

The overall spectral shape of a red giant as warm as Arcturus is
primarily determined by continuous opacities and metals, but also
by a small contribution of CO. Water vapor is, for example,  not
expected in the spectrum of Arcturus. This is in stark contrast to
cooler giants where molecules (such as TiO, VO, H$_2$O, OH, CO,
and SiO) are totally dominant, the metals only playing a minor
role. The main continuous opacity source in the optical is due to
H$^{-}_{\rm{bound-free}}$, i.e. ionization and recombination of
the H$^-$-ion. Further out in the infrared, H$^-_{\rm{free-free}}$
opacity (i.e. bremsstrahlung and absorption by an electron in the
field of a neutral hydrogen atom) is most important. This opacity
source grows with wavelength and dominates beyond 1.6 microns. At
1.6 $\mu$m there is a minimum in the opacity, which means that the
continuum at 1.6 microns is formed relatively far down in the
photosphere. As we go further out in the infrared the continuum is
formed successively further up in the photosphere. We therefore
expect also mid-infrared lines to be formed further out than lines
at, say, 1.6 microns.

\section{PROCEDURE}

 We start off by testing our model photosphere and the stellar
 parameters used by calculating a synthetic spectrum of the
 1.6 micron region, and confronting it with the observed,
 high-resolution, high-signal-to-noise spectrum
 found in the Arcturus Atlas (Hinkle et al. 1995).
 The OH first overtone lines found in this region are, as was noted
 above, formed at a deep level in the photosphere where
 the model assumptions are certainly valid. We are able to
 model the OH lines well and are thus confident that
 our parameters are correct.

 Next, we turn our attention to the 12 micron region to study the
 pure rotational lines. The continuum is formed further out
 at these wavelengths and the cores of moderately strong lines
 could be formed in shallow layers in the atmosphere where the
 model assumptions may be erroneous.
 We find that the OH($v=1-1$) and OH($v=2-2$) lines can be modelled well.
However, the model cannot account for the deep cores of the
OH($v=0-0$) lines. Furthermore, the model does not predict any
water lines at all, which is a large fallacy.

Since we are working with high-resolution data, our lines are
resolved, thus providing kinematic information. Based on this
piece of information, the line widths, and the excitation
temperatures of the lines, we argue that all the lines are formed
within the photosphere, and not above the photosphere, for example
in a circumstellar envelope (see Ryde et al. 2002).

We therefore construct a semi-empirical model-atmosphere based on
our {\sc marcs} model, by adjusting the temperature profile in the
outer most, uncertain, layers of the model photosphere. The model
now violates the condition of flux constancy imposed on {\sc
marcs} models. Electron and gas pressures were recalculated
assuming hydrostatic equilibrium. The density and standard opacity
are also consistently recalculated from this new temperature
structure, as are the ionization balance, molecular equilibria,
and opacities. New synthetic spectra were calculated assuming, as
before, LTE for the molecular equilibrium and the line formation.
The revisions to the upper boundary layer made by us
were made solely with the aim of fitting observed line profiles.

We are successful in finding a semi-empirical model that predicts
the strengths of all the OH lines, the water vapor lines, and also
the CO vibration-rotation, 4.6 micron lines, the cores of which
are also formed far up in the photosphere. Over most of the
modified region, the temperatures are a mere 300 K cooler than the
original {\sc marcs} model. In Figure 1 a few water vapor lines
are shown together with the synthetic spectrum based on the
original {\sc marcs} atmosphere and the spectrum based on the
semi-empirical model photosphere.

\begin{figure}
\plotone{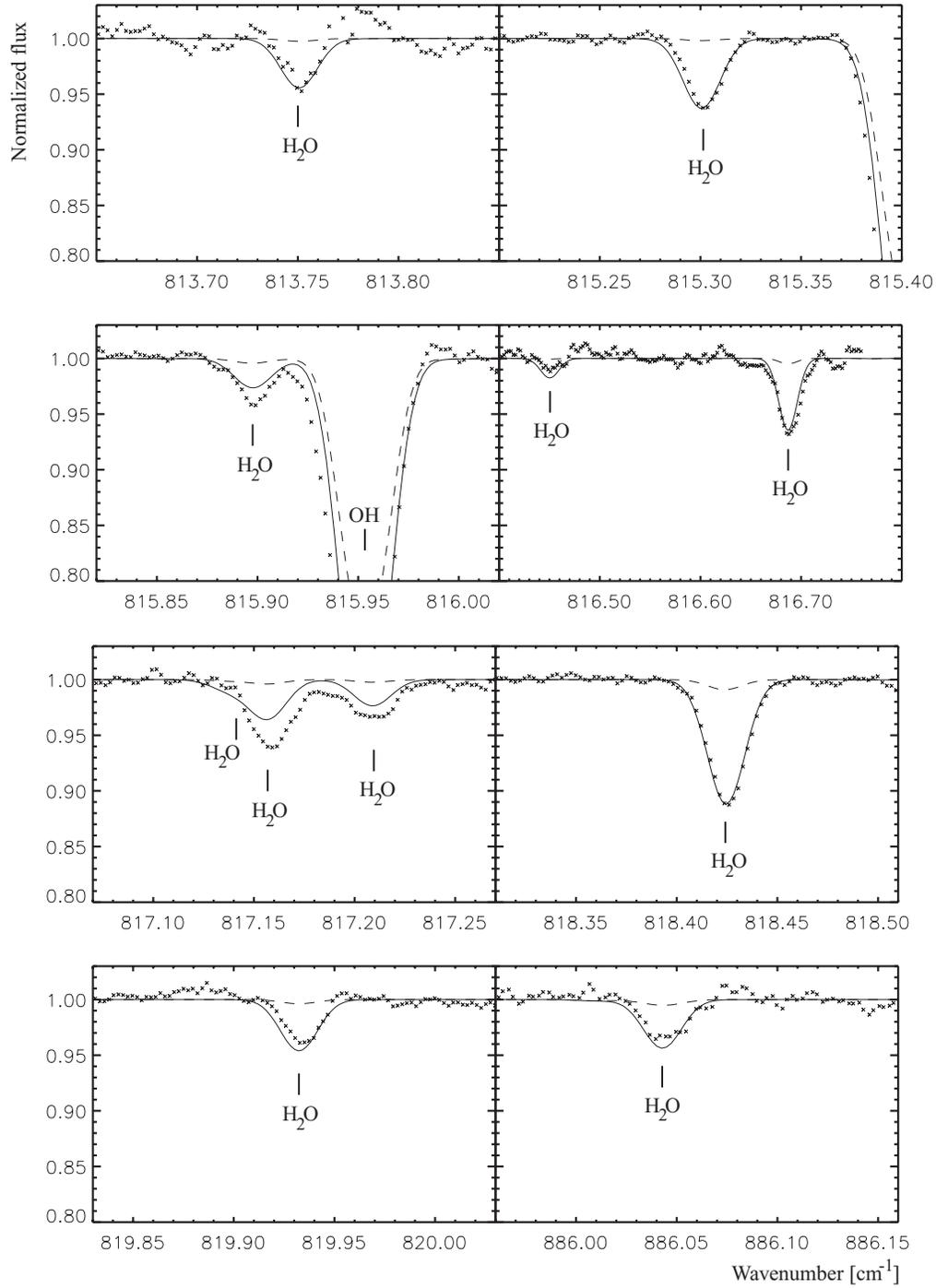}  \caption{ This figure of pure rotation,
water-vapor lines is taken from Ryde et al. (2002). The observed
spectrum is plotted with crosses. The dashed line shows the
synthetic spectrum calculated based on our original model
photosphere. The full line shows the spectrum based on the
photosphere with the new temperature structure. }
\end{figure}

\section{DISCUSSION}

What could cause the detected stellar steam in the disk-averaged
spectrum of this K2 giant? The outer layers of the atmosphere seem
to be more complicated than the simple assumptions made in our
models. What Physics are we missing? We will now discuss a few
points related to this.

First, relaxation of the  assumptions behind the {\sc marcs}
models may change the atmospheric structure. Even small
alterations of the heating or cooling terms in the energy equation
(for example due to dynamic processes or uncertainties and errors
in the calculations of radiative cooling) may lead to changes in
the temperature structure in the outer, tenuous regions of the
photosphere where the heat capacity per volume is low. The
assumption of LTE is one target of model builders seeking to go
beyond the standard assumptions. Atomic -- continuous and line --
opacity was treated in non-LTE using large model atoms by Short
(2002) in spherical geometry, hydrostatic equilibrium, and flux
constancy. Introduction of non-LTE in place of LTE lowered the
boundary temperature by an amount quite similar to our empirical
cooling of the {\sc marcs} model. This suggests that non-LTE
effects in atoms may suffice to account for the surprising
presence of H$_2$O molecules in the photosphere.

Second, there are certainly inhomogeneities over the surface of
the star. A key assumption made in construction and application of
the model atmosphere is that physical conditions are everywhere
the same at a given geometrical depth. Such a model is said to be
`homogeneous'. Arcturus, with its deep convective envelope, is
certain to be inhomogeneous; it is likely to exhibit surface
granulation which may be crudely characterized as consisting of
columns of warm rising and cool sinking gas. The continuum
intensity contrast in the mid-infrared will be small. There will,
however, most likely be a sharp difference in the H$_2$O column
density with the sinking column much richer in the molecules, and,
as a result, the surface-averaged spectrum will show stronger
H$_2$O lines than the spectrum of the equivalent homogeneous model
atmosphere. Unfortunately,  inhomogeneous model atmospheres of
giant stars are unavailable.
Yet, the modest 300 K cooling imposed on the {\sc marcs} model is
surely within the range of temperature differences between rising
and sinking granules.

An alternative mechanism for producing inhomogeneities is
variously known as `a molecular catastrophe' or `temperature
bifurcation'.
 In warm stars, the CO
molecular opacity influences the temperature profile predicted by
a program like {\sc marcs}.
 If the temperature of gas in the boundary layer
is perturbed, there may be a runaway to one of two different
solutions. Suppose the gas is cooled, more CO molecules form and
increase the cooling of
 the
boundary layer. The temperature drop leads to yet more CO
molecules and to a runaway to a `cool' boundary layer. Conversely,
if the perturbation raises the temperature, CO molecules are
dissociated, the cooling rate  decreased, and the temperature rise
is continued.  In cool O-rich stars, where the CO number density
is limited by the abundance of carbon, this sensitivity of opacity
to temperature is greatly reduced, but other molecules (e.g., SiO
or H$_2$O) may act in a similar way.

Third, Arcturus certainly exhibits a chromosphere, which has been
modelled by Ayres \& Linsky (1975). The chromospheric temperature
rise starts at approximately $\log\tau_{500}=-2.7$ which is deeper
down in the photosphere than the on-set of our suggested
temperature decline. This bifurcated temperature structure has
been found earlier for Arcturus based on CO 4.6 $\mu$m data
(Wiedemann et al. (1994)) and has also been discussed for the Sun
(see, for example, Ayres 2002). Existence of the chromosphere
around Arcturus suggests a heating mechanism active in and above
the upper photosphere.  Cuntz \& Muchmore (1989) made a
preliminary study of the effects of propagating acoustic waves
including molecular line cooling. When the waves lead to weak
shocks, a cool atmosphere is formed. On the other hand, strong
shocks lead to a temperature inversion, approximately matching an
empirical chromospheric model for Arcturus. These calculations
incorporating a driving mechanism for temperature bifurcation
support the idea of an inhomogeneous upper photosphere and low
chromosphere.

Chromospheres are indeed dynamic, and are thought to be a
wave-driven phenomenon showing spatially and temporally
intermittent structures, see for example the work on the solar
chromosphere by Carlsson \& Stein (1995). They find that the mean
temperature is low, but also that transient, high temperatures
give rise to the observed enhance emission in the ultra-violet. In
the 3D simulations of the solar chromosphere by Wedemeyer et al.
(2002), accoustic shocks are excited by the solar convection zone
and form structures with different temperatures caused by
interaction of propagating wave fronts. They show that the cool
phases lasts for approximately 1 minute, on average, until a new,
hot wave front passes by again. Whether the timescales are long
enough and the physical and chemical conditions are right for
molecular formation and how molecules react to the shocks is not
clear. In this context it is interesting to note that our
suggested model that accounts for the water-vapor, OH, and CO
features also predicts strong TiO features in the optical, which
is not observed. The abundance of titanium is more than three
orders of magnitude lower than that of oxygen and carbon, perhaps
suggesting that TiO may not have time to form between shocks.

\section{CONCLUDING REMARKS}



I have discussed our unexpected discovery of absorption lines of
water vapor in the disk-averaged spectrum of Arcturus (Ryde et al.
2002). Based on kinematic information, line widths and excitation
temperature, we argue that the water vapor is photospheric and not
circumstellar. Until now, H$_2$O has not been expected to exist in
photospheres of stars of Arcturus' effective temperature. This is
true unless the outer layers of the model photosphere do not
describe these regions properly. By modifying these layers we have
succeeded in finding {\it a} model which yields a synthetic
spectrum matching the observations. We are, however, not claiming
that this necessarily is the only and true model. Our exercise
simply shows that it is possible to achieve a photospheric
spectrum containing water vapor also for early K giants if the
outer parts are cooled for some reason. Possible reasons for the
unexpected water vapor include photospheric inhomogeneities (such
as convective flows, molecular catastrophes, or even star spots)
and departures from LTE in the photospheric structure or line
formation in the boundary layers. 3D hydrodynamic models of Red
giants do not as yet exist, but may be a crucial tool for the
understanding of how water vapor can exist on the surface of a
warm star such as Arcturus. We strongly encourage model makers to
pursue the difficult task of modelling red giants
hydrodynamically.

\section{ACKNOWLEDGEMENTS}
This work was supported by the Swedish Foundation for
International Cooperation in Research and Higher Education,
Stiftelsen Blanceflor Boncompagni-Ludovisi, n\'ee Bildt, and the
Robert A. Welch Foundation of Houston, Texas. The construction of
the {\sc texes} spectrograph was supported by grants from the {\sc
nsf}, and observing with {\sc texes} was supported by {\sc usra}
the Texas Advanced Research Program. NSO/Kitt Peak FTS data used
here were produced by NSF/NOAO. We acknowledge the support of the
{\sc irtf} staff.

\end{document}